\documentclass[final,3p,times]{elsarticle}

\usepackage{amssymb}
\usepackage{amsmath}
\usepackage{bm}

\usepackage{xcolor}
\usepackage[normalem]{ulem}

\journal{Journal of Subatomic Particles and Cosmology}

\begin{document}

\begin{frontmatter}



\title{Probing the QCD Phase Structure with Dileptons from SIS to LHC Energies}

\author[ddd,aaa,bbb]{Adrian William Romero Jorge}
\author[ccc]{Taesoo Song} 
\author[fff]{Qi Zhou}
\author[ccc,aaa,bbb]{Elena Bratkovskaya}

 \affiliation[ddd]{organization={Frankfurt Institute for Advanced Studies (FIAS)},
             addressline={Ruth Moufang Str. 1},
             city={Frankfurt},
             postcode={60438},
             country={Germany}} 
             
\affiliation[aaa]{organization={Helmholtz Research Academy Hessen for FAIR (HFHF)},
             addressline={GSI Helmholtz Center for Heavy Ion Physics. Campus Frankfurt},
             city={Frankfurt am Main},
             postcode={60438},
             country={Germany}}

 \affiliation[bbb]{organization={Institute for Theoretical Physics, Johann Wolfgang Goethe University},
             addressline={Max-von-Laue-Str. 1},
             city={Frankfurt am Main},
             postcode={60438},
             country={Germany}}
                       
 \affiliation[ccc]{organization={GSI Helmholtzzentrum fur Schwerionenforschung GmbH},
             addressline={Planckstr. 1},
             city={Darmstadt},
             postcode={64291},
             country={Germany}}

\affiliation[fff]{organization={State Key Laboratory of Nuclear Physics and Technology, Institute of Quantum Matter, South China Normal University},
city={Guangzhou},
postcode={510006},
country={China}}

\begin{abstract}
We study the properties of strongly interacting matter at finite temperature and baryon chemical potential in relativistic heavy-ion collisions, with emphasis on dilepton probes of the QCD phase structure. The equilibrium QGP is described within the Dynamical QuasiParticle Model (DQPM), which reproduces the lattice-QCD equation of state and provides $\mu_B$-dependent quasiparticle properties, transport coefficients, and thermal dilepton rates, including elastic and inelastic partonic processes. The dynamical evolution is modeled with the off-shell Parton--Hadron--String Dynamics (PHSD) transport approach, which consistently propagates partonic and hadronic degrees of freedom and incorporates chiral-symmetry restoration effects. We discuss the space--time evolution of heavy-ion collisions over a broad energy range and show that a small deconfined QGP core can already emerge at $\sqrt{s_{NN}}\simeq 3.5$ GeV. We present, for the first time, the baryon-chemical-potential dependence of QGP thermal dilepton radiation in PHSD and demonstrate that its influence increases toward lower collision energies. The excitation function indicates that thermal QGP radiation can exceed dileptons from correlated charm decays in central Au+Au collisions at $\sqrt{s_{NN}}\lesssim 25$--$30$ GeV, making RHIC--BES and FAIR energies particularly promising for the direct observation of QGP electromagnetic radiation after subtraction of heavy-flavor and Drell--Yan contributions.
\end{abstract}




\end{frontmatter}


We explore the properties of strongly interacting matter at finite temperature and baryon chemical potential as created in relativistic heavy-ion collisions, focusing on the QCD phase structure probed via dilepton observables. The equilibrium description of the non-perturbative quark–gluon plasma (QGP) is realized within the Dynamical QuasiParticle Model (DQPM), which reproduces lattice QCD results for the equation-of-state (EoS) above the deconfinement temperature. 
The dynamical, non-equilibrium evolution of strongly interacting matter is described within the off-shell transport approach Parton–Hadron–String Dynamics (PHSD) \cite{Cassing:2008sv,Moreau:2019vhw}, which consistently propagates partonic and hadronic degrees-of-freedom and their interactions based on Kadanoff–Baym theory. This allows for a microscopic realization of the QCD phase transition while maintaining energy–momentum and quantum number conservation. Chiral symmetry restoration effects are included, leading to in-medium modifications of hadronic spectral functions and strange hadron properties.
The non-perturbative QGP is described in  the DQPM in terms of massive quarks and gluons with finite spectral widths, formulated in a two-particle irreducible (2PI) propagator representation. Its temperature $T$  and baryon chemical-potential $\mu_B$ dependent quasiparticle properties are fitted to reproduce the lattice-QCD equation of state at $\mu_B=0$, providing effective interactions and partonic cross sections for transport calculations in the whole $(T,\mu_B)$ plane.

Key dilepton sources in PHSD include hadronic decays, bremsstrahlung,  QGP radiation ($q+\bar q \to e^+e^-$, \ $q+\bar q \to g+ e^+e^-$, \ $q+g \to q+ e^+e^-$), primary Drell-Yan production, and semileptonic decays of correlated charm and bottom pairs. 
As shown in  Ref. \cite{Jorge:2025wwp},  PHSD well describes dilepton data from HADES, STAR, and ALICE experiments. 
In this contribution we highlight the results from this study.


In Ref.~\cite{Jorge:2025wwp}, we investigated the $\mu_B$ dependence of QGP radiation within PHSD using dilepton rates calculated in the DQPM. The effect of finite $\mu_B$ on the QGP dilepton yield becomes more pronounced at lower collision energies, where larger baryon chemical potentials are reached. However, its impact on the total dilepton spectra remains small because the space-time volume of the QGP decreases rapidly with decreasing collision energy.

\begin{figure}[t!]
    \centering
\includegraphics[width=0.329\linewidth]{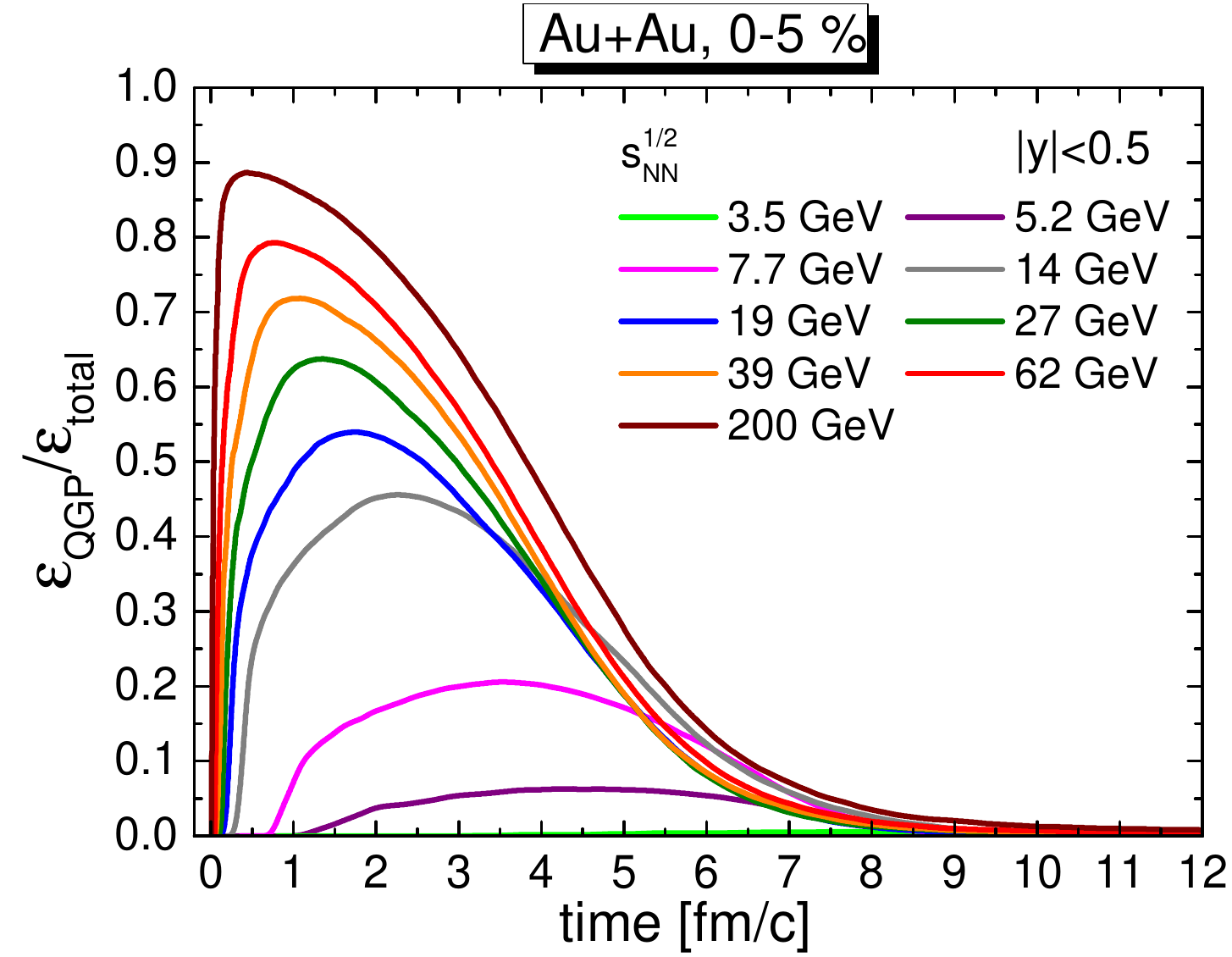}
\includegraphics[width=0.329\linewidth]{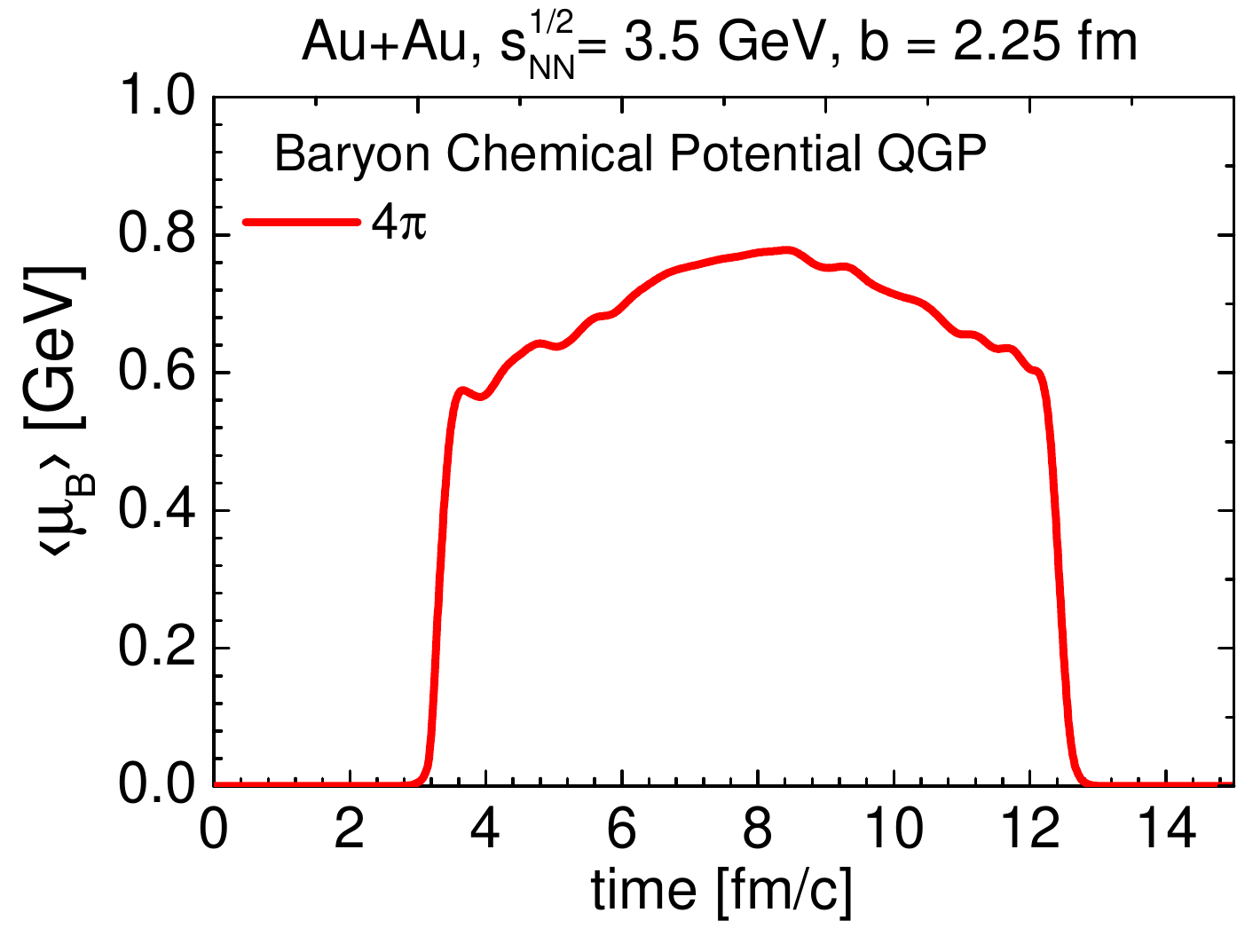} 
\includegraphics[width=0.329\linewidth]{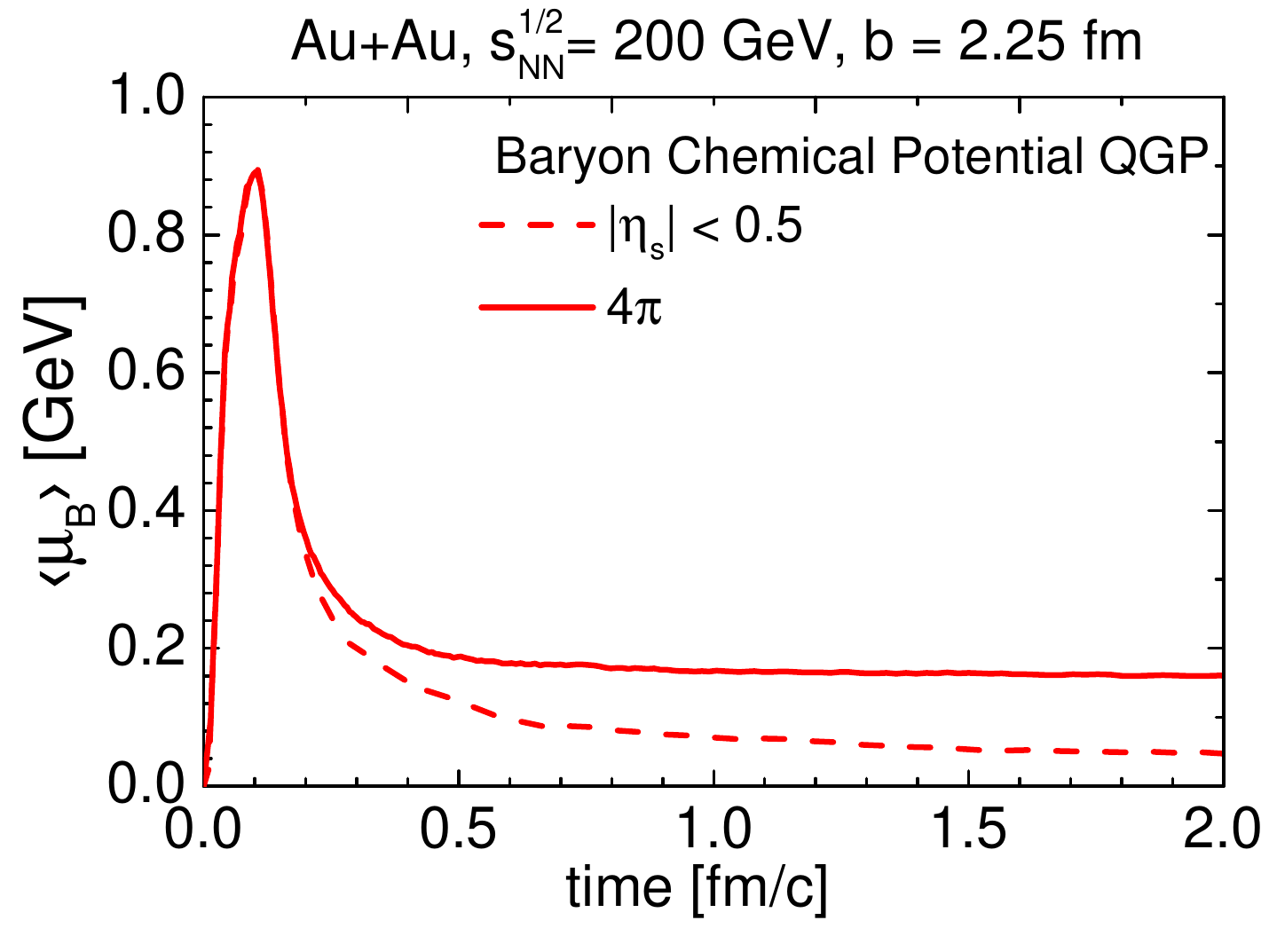}
\caption{Left: The fraction of energy in the QGP phase as a function of time $t$ for central Au+Au collisions (impact parameter $b = 2.25$ fm) at various collision energies $\sqrt{s_{NN}}$ from 3.5 to 200 GeV at the midrapidity  $|y| < 0.5$. 
Middle, right: Averaged  baryon chemical potential - as probed by the dileptons from QGP -  versus time for central ($b=2.25$ fm) Au+Au collisions at $\sqrt{s_{NN}}=3.5 $ GeV (middle) and 200 GeV (right). The solid lines correspond to the full space while the dashed line stand for the spatial rapidity $|\eta_s|<0.5$.
   }
    \label{Fraction_QGP}
\end{figure}

To quantify this effect, Fig.~\ref{Fraction_QGP}~(left) shows the time evolution of the fraction of energy contained in the QGP phase for central Au+Au collisions, $b=2.25$~fm, corresponding approximately to 0--5\% centrality, at collision energies $\sqrt{s_{NN}}=3.5$--200~GeV within the midrapidity interval $|y|<0.5$. At high energies, the early-time evolution is dominated by partonic matter; at $\sqrt{s_{NN}}=200$~GeV, up to about 90\% of the midrapidity energy is carried by the QGP. This fraction decreases strongly with decreasing collision energy. Nevertheless, even at the lowest energy, $\sqrt{s_{NN}}=3.5$~GeV, a very small QGP component, below 1\% of the total energy, is still formed.
The middle and right panels of Fig.~\ref{Fraction_QGP} show the time dependence of the average baryon chemical potential probed by QGP dileptons in central Au+Au collisions at $\sqrt{s_{NN}}=3.5$ and 200~GeV, respectively. Solid lines correspond to the full space-time volume, while dashed line indicate the spatial rapidity interval $|\eta_s|<0.5$. At $\sqrt{s_{NN}}=3.5$~GeV, partonic dileptons probe matter only slightly above the critical energy density $\varepsilon_C$ but at large baryon chemical potential, $\mu_B \gtrsim 0.6$~GeV. In contrast, at $\sqrt{s_{NN}}=200$~GeV, $\mu_B$ rapidly decreases toward nearly zero values at midrapidity.

\begin{figure}[h!]
    \centering
\includegraphics[width=0.30\linewidth]{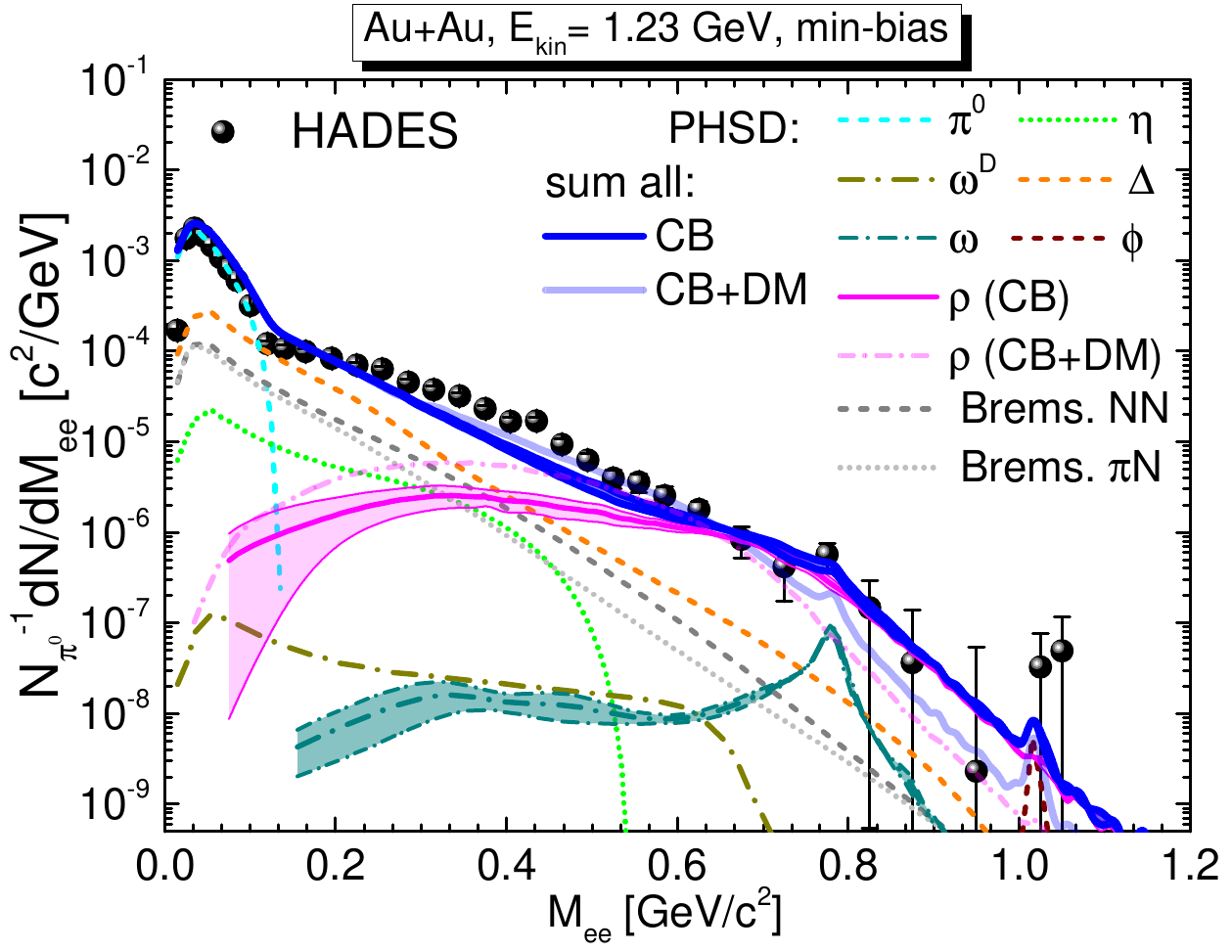}
\includegraphics[width=0.32\linewidth]{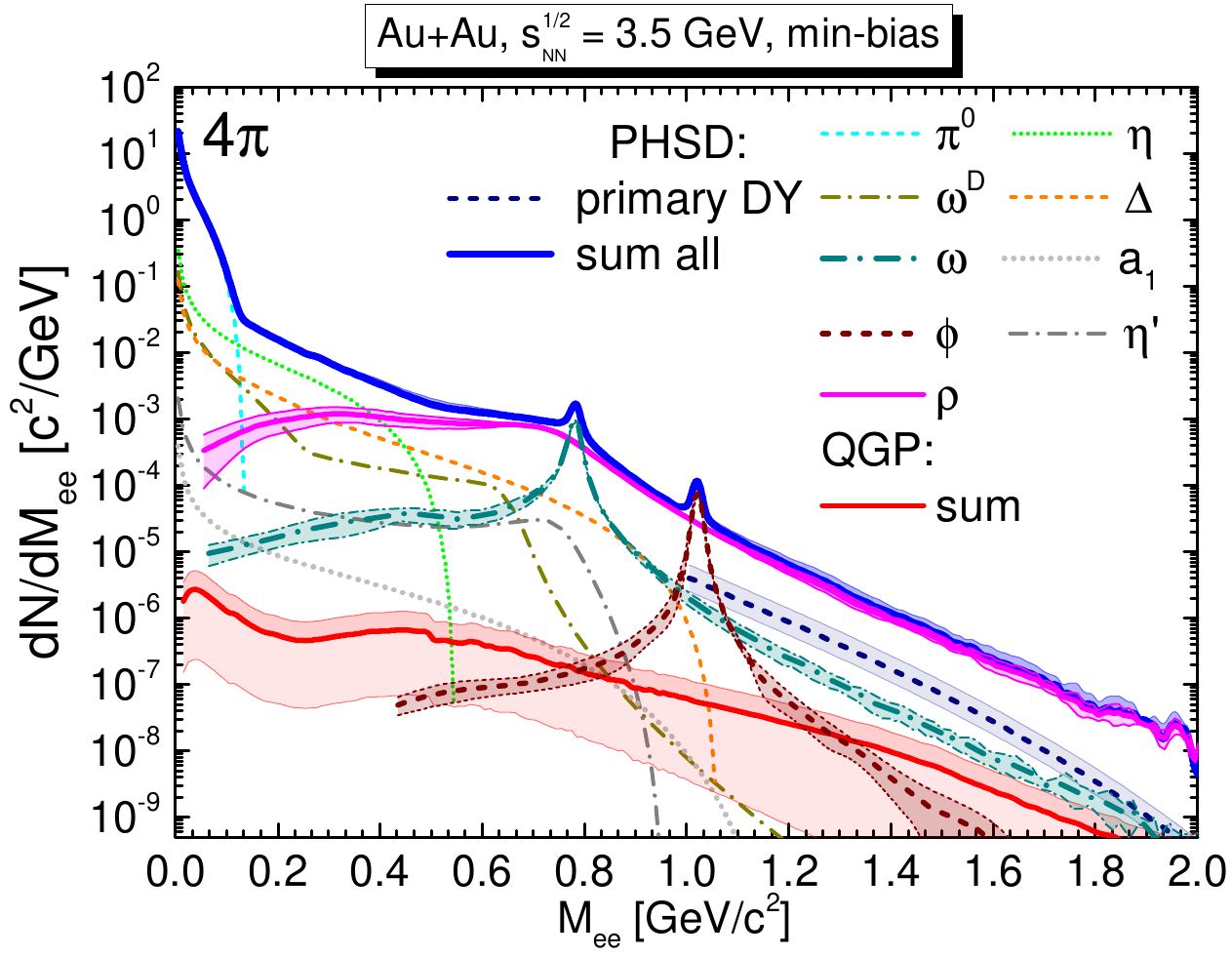}
\includegraphics[width=0.32\linewidth]{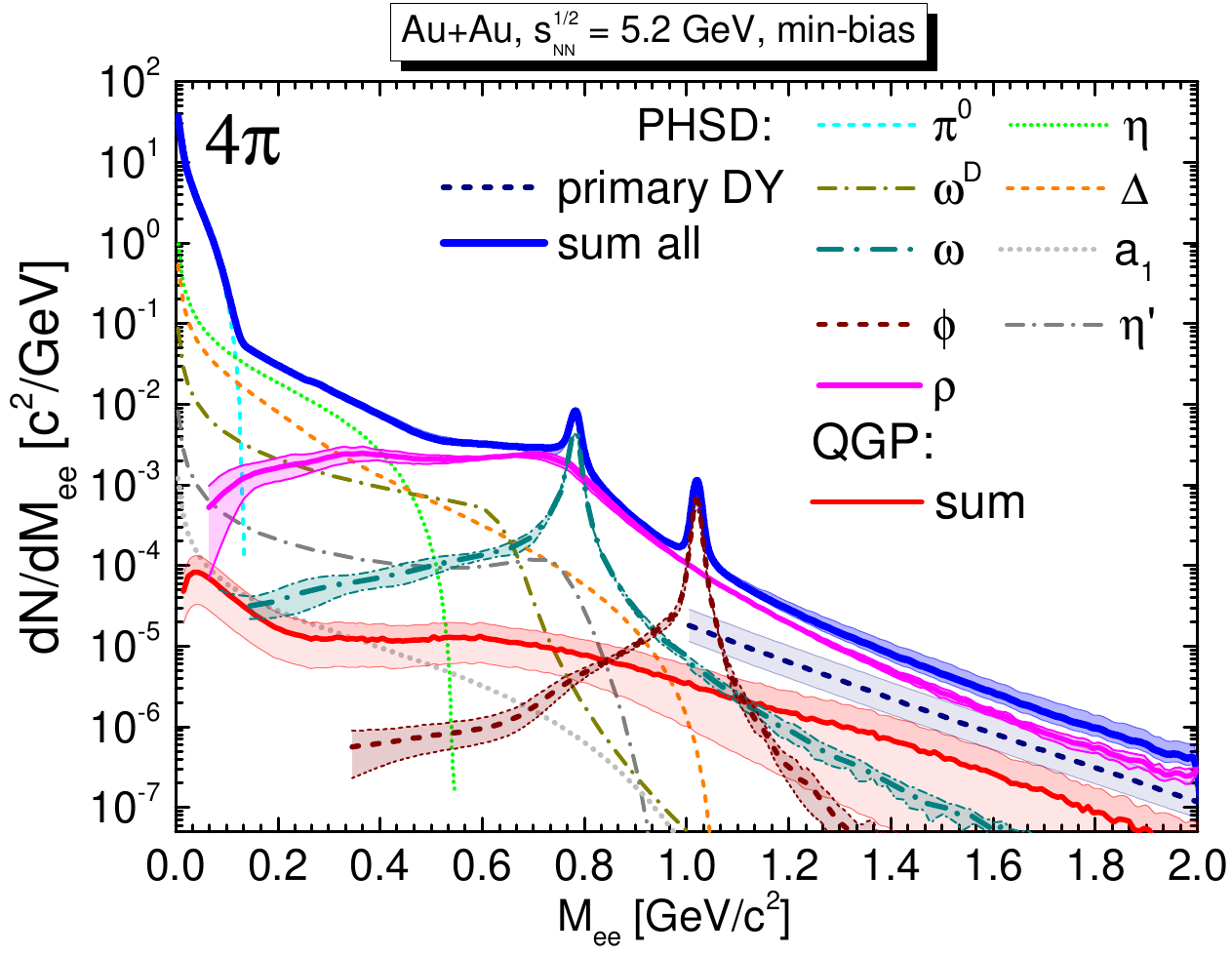}
    \caption{Left: Invariant mass spectra of dileptons $dN/dM_{ee}$ — normalized to the $\pi^0$ multiplicity - from PHSD for Au + Au collisions at 1.23 AGeV in comparison to the experimental measurements by the HADES collaboration \cite{HADES:2019auv}.
    Middle, right:  The PHSD predictions for the invariant mass spectra $dN/dM_{ee}$ of dileptons for Au + Au collisions for minimal bias at  $\sqrt{s_{NN}}=$3.5  GeV (middle) and  5.2  GeV (right).  
      The total yield is displayed in terms of the blue lines while the different contributions are specified in the legends.   
 }
    \label{mass_spectra_SIS18}
\end{figure}
\begin{figure}[h!]
    \centering
\includegraphics[width=0.32\linewidth]{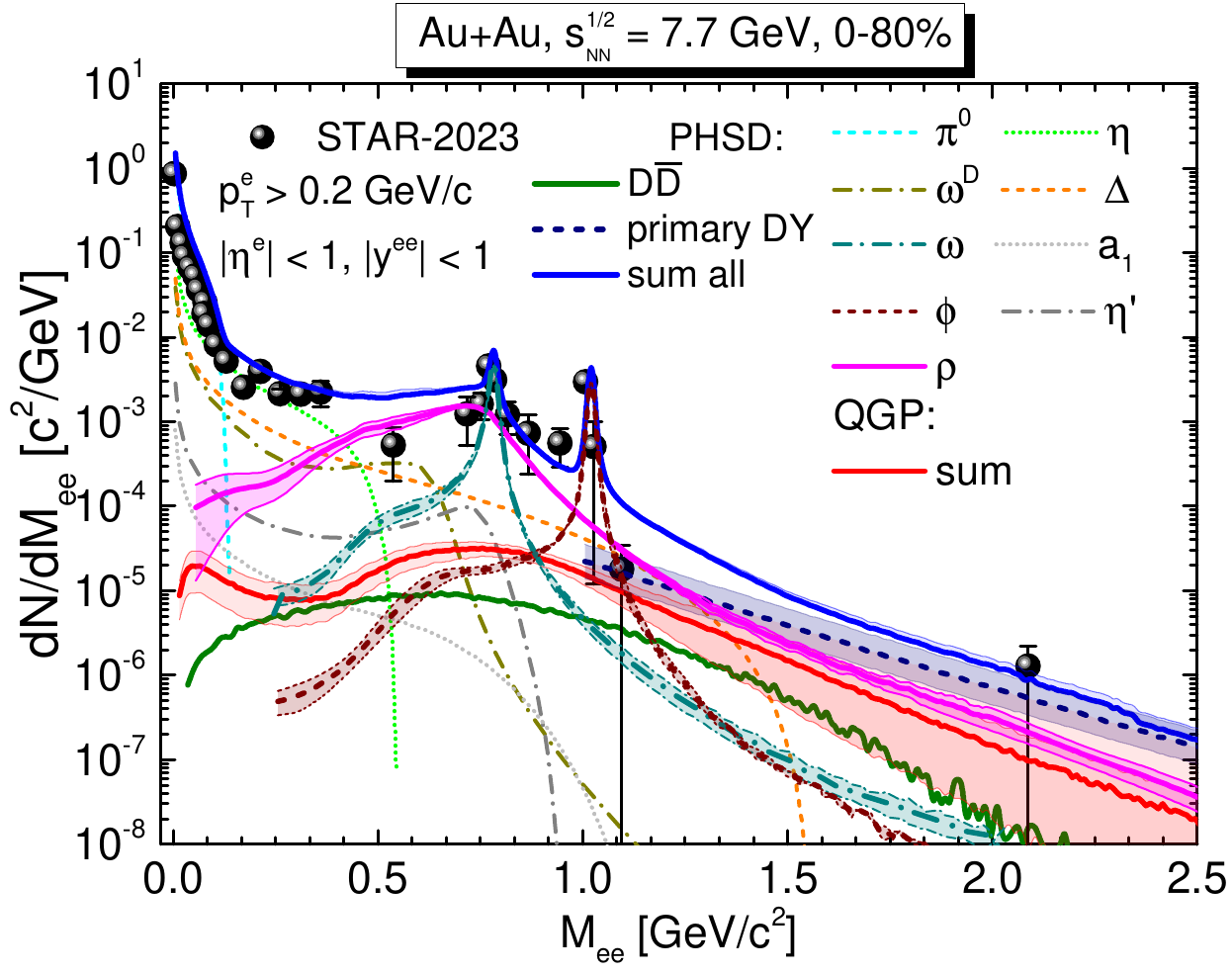}
\includegraphics[width=0.32\linewidth]{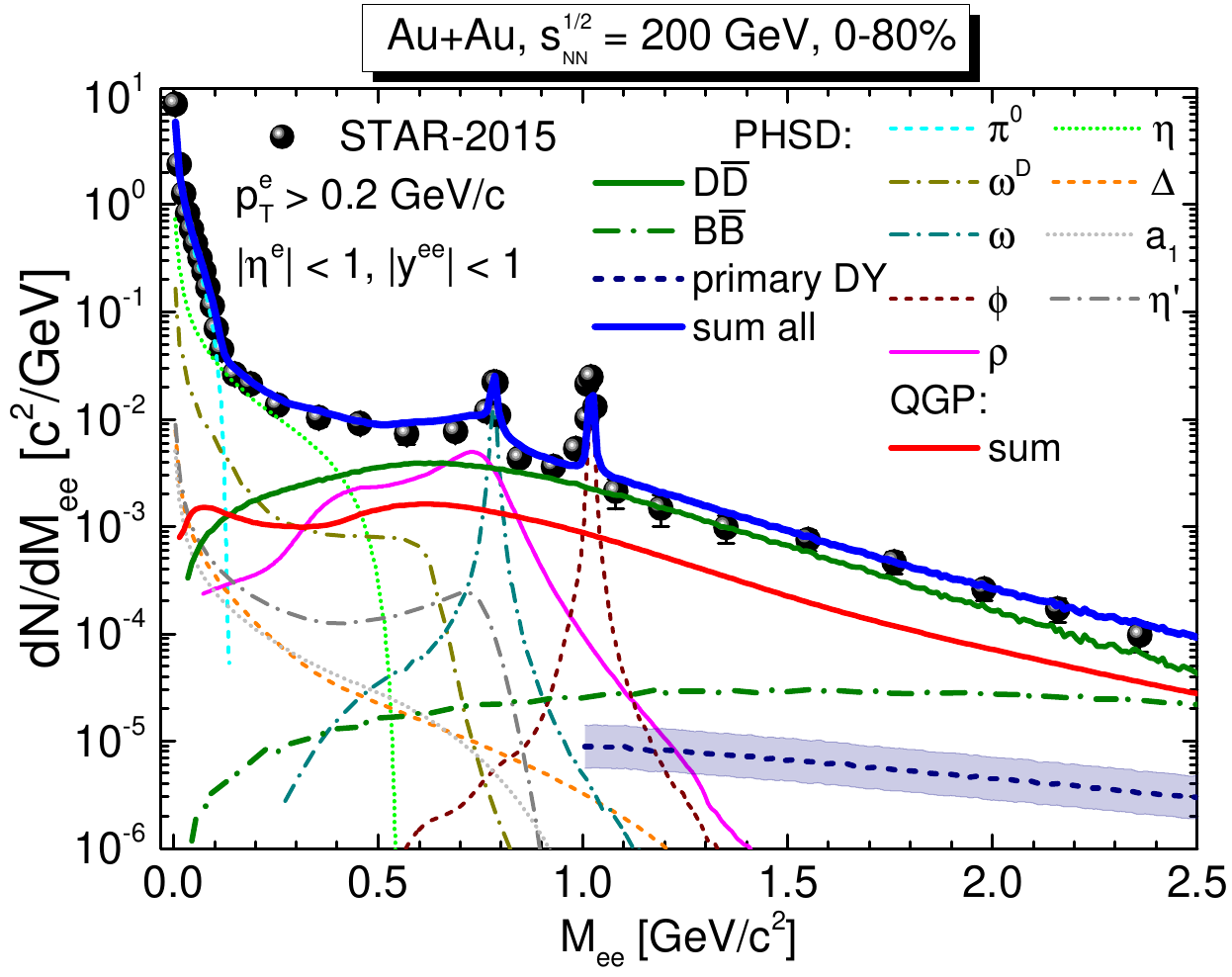}
\includegraphics[width=0.32\linewidth]{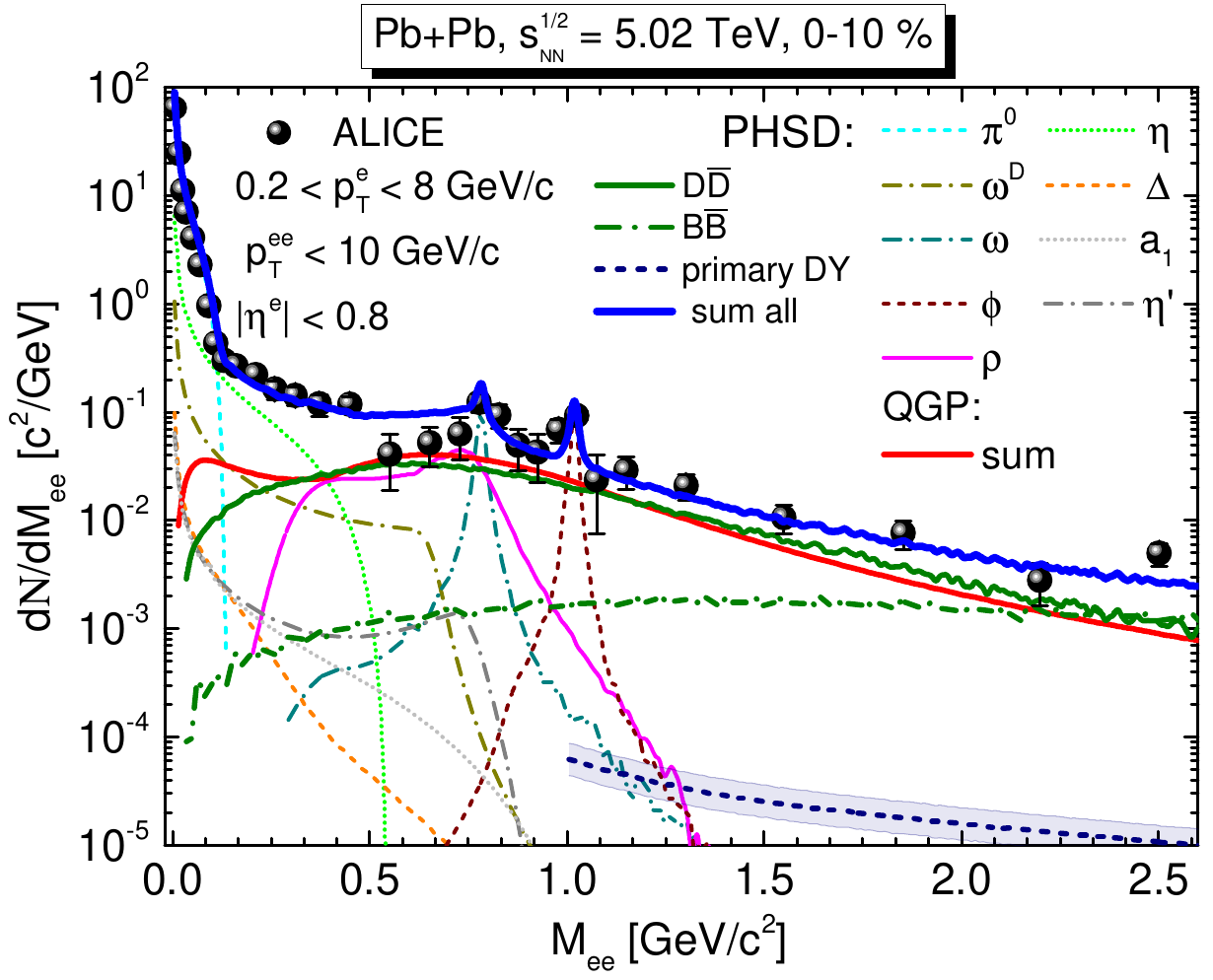} 
    \caption{Left, middle: Invariant mass spectra of dileptons from PHSD in comparison to STAR data \cite{STAR:2023wta,Han:2024nzr,STAR:2015tnn}  for Au+Au collisions for $0-80$\% centrality at $\sqrt{s_{NN}}=7.7$ and 200 GeV.
 Right: Invariant mass spectra of dileptons from PHSD  $dN/dM_{ee}$  for Pb+Pb
 5.02 TeV  for 0-10\% central  collisions in comparison to the experimental measurements by the ALICE collaboration \cite{ALICE:2018ael,ALICE:2023jef,Gunji:2017kot,Meninno:2020mjd}.    
    }
    \label{mass_spectra_RHIC}
\end{figure}
\begin{figure}[h]
    \centering
\includegraphics[width=0.41\linewidth]{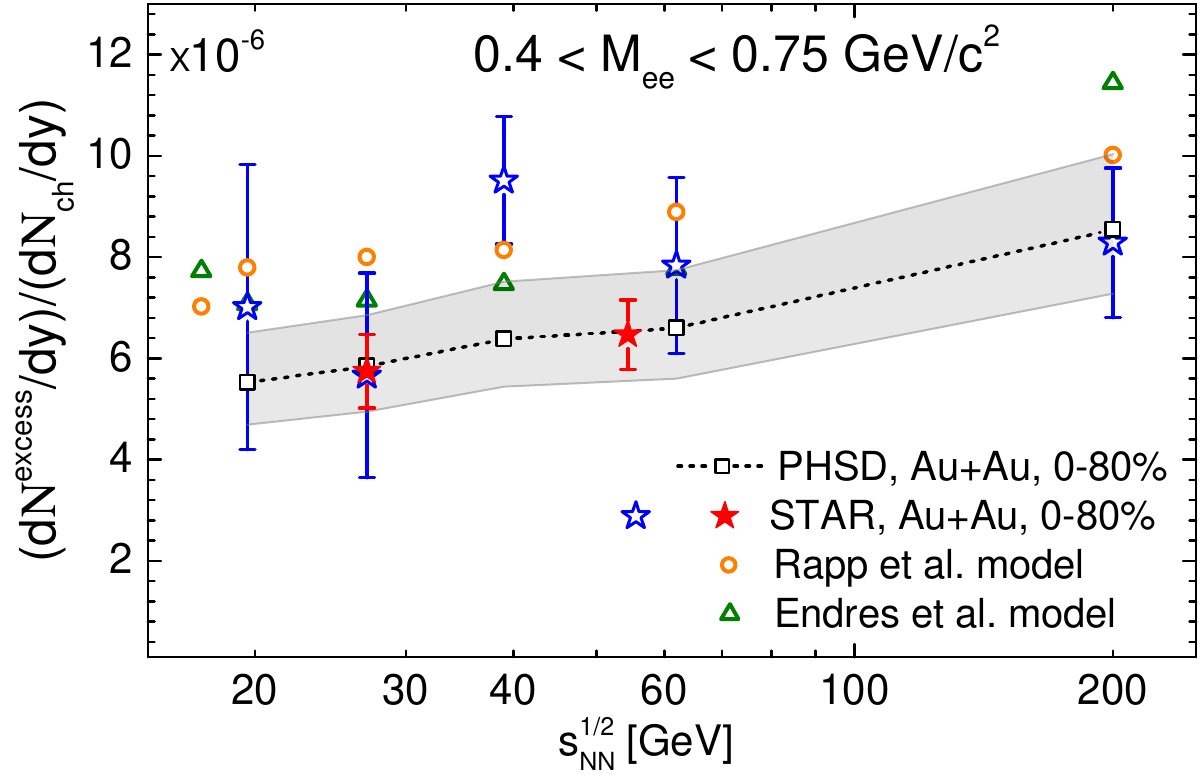} \hspace*{5mm}
\includegraphics[width=0.37\linewidth]{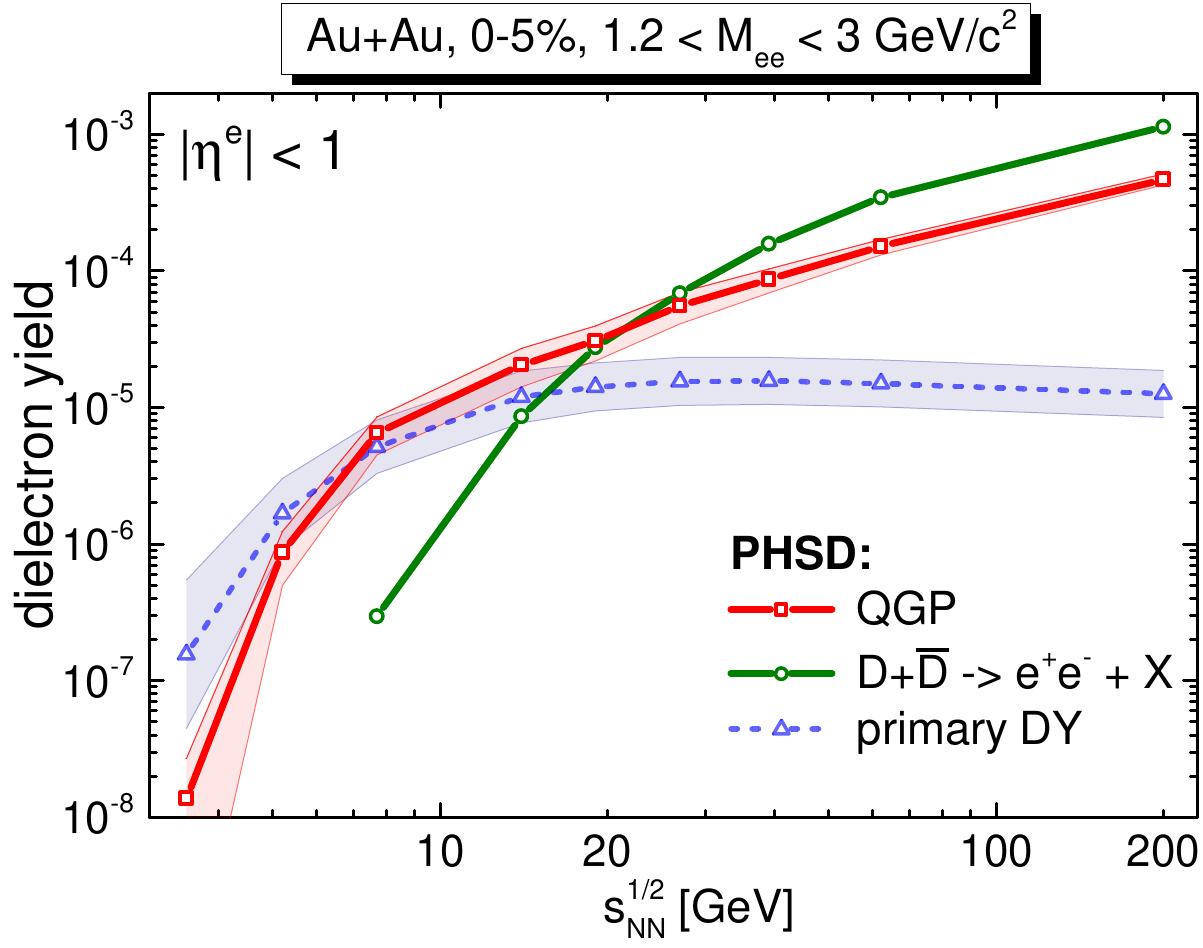}
    \caption{Left: Collision energy dependence of the integrated dilepton excess yields for $0.4 < M_{ee} < 0.75$ GeV/c$^2$, normalized by $dN_{ch}/dy$ \cite{STAR:2023wta}, for 0-80 \% central Au + Au collisions. Black open squares (connected by the black dashed line) represent the  PHSD results including theoretical uncertainties in centrality determination for $dN_{ch}/dy$ shown by a shaded gray.  The STAR acceptance has been applied. 
    The experimental data from  the STAR Collaboration  are shown by the blue stars \cite{STAR:2023wta} and red stars \cite{STAR:2024bpc}. 
    The theoretical calculations from Rapp et al.  (orange circles) \cite{Rapp:2014hha} and Endres et al. (olive triangles) \cite{Endres:2016tkg} are also presented for comparison.  
    Right: Comparison of the dielectron yield from the $D\bar{D}$ (green), primary Drell-Yan (DY) (dashed blue) and  QGP (red) contributions integrated in the interval $1.2 < M_{ee} < 3$ GeV/c$^2$ as a function of the center of mass energy $\sqrt{s_{NN}}$ from 3.5 GeV to 200 GeV, for  Au + Au collision 0-5\%.     
    }
    \label{excess_yield}
\end{figure}

In Figures \ref{mass_spectra_SIS18} and \ref{mass_spectra_RHIC} we show the dileptons spectra from SIS to RHIC and LHC energies. One can see a good agreement of the PHSD results with experimental data from HADES, STAR and ALICE collaborations.
Moreover, the middle and right plots in Fig. \ref{mass_spectra_SIS18} show the PHSD predictions for the future CBM experiment at FAIR (Darmstadt).
One can see that at SIS energies (1–2\,$A$GeV), hadronic processes such as bremsstrahlung and $\pi^0$, $\eta$ and $\Delta$ Dalitz decays dominate the dilepton yield in the low-mass region, whereas the $\rho$ meson broadening plays a decisive role in reproducing the observed spectral shape.  The enhancement of the dilepton yield in A+A compared to p+p collisions (which grow with the system size $A$) - as observed by the HADES collaboration - is attributed to  in-medium effects such as multi-particle reactions - as $\Delta$ regeneration which leads to enhanced dilepton radiation as well as the issue of partial chiral symmetry restoration manifested in the collisional broadening of the vector meson spectral functions.
The  vector meson contribution grows with increasing energy and dominates the spectra at low mass region $M_{ee}<0.7$ GeV/c$^2$. 
With increasing energy the QGP contribution grows and dominates the intermediate mass spectra 
$1.2 < M_{ee} < 3$ GeV/c$^2$ together with dileptons from correlated charm ($D\bar{D}$).
Moreover, the primary DY processes contribute visibly at intermediate energies to  the same mass region.

To quantify the relative importance of different dilepton sources, Fig.~\ref{excess_yield} shows the excitation functions of integrated dilepton yields. The left panel presents the excess yield in the low-mass region, $0.4<M_{ee}<0.75$~GeV/$c^2$, normalized by $dN_{\rm ch}/dy$~\cite{STAR:2023wta}, for 0--80\% central Au+Au collisions. The right panel shows the contributions from correlated $D\bar{D}$ decays, primary Drell--Yan (DY) production, and QGP radiation in the intermediate-mass region, $1.2<M_{ee}<3$~GeV/$c^2$, as a function of $\sqrt{s_{NN}}$ from 3.5 to 200~GeV.
The low-mass excess is well described by the collisional broadening of the vector-meson spectral functions, in particular those of the $\rho$, $\omega$, and $\phi$ mesons. This mechanism reproduces the dilepton enhancement over the hadronic cocktail observed at RHIC BES energies, as well as the STAR excitation function of the integrated excess yield in the mass range $0.4<M_{ee}<0.75$~GeV/$c^2$. In the intermediate-mass region, the dominant sources are partonic radiation and correlated heavy-flavor decays. The calculated excitation function shows that the QGP contribution exceeds the correlated charm contribution at $\sqrt{s_{NN}}\sim 30$~GeV in central collisions and at $\sqrt{s_{NN}}\sim 20$~GeV for minimum-bias collisions, in line with Ref.~\cite{Song:2018xca}. Although correlated charm remains the dominant dilepton source at LHC energies, the QGP contribution is also sizable; after subtraction of the charm component, intermediate-mass dileptons provide direct access to very hot QGP matter at $\mu_B\simeq 0$.
We also find that primary Drell--Yan production can contribute significantly at intermediate collision energies. Its calculation, however, is affected by sizable uncertainties, especially in the limited phase space available at these energies, and is sensitive to the employed parton distribution functions. Therefore, the present DY estimates should be regarded as upper estimate.
Precise dilepton measurements in $p+p$ collisions are essential for constraining and subtracting the DY contribution from A+A spectra, thereby improving the extraction of QGP dilepton radiation.

Future measurements at FAIR, NICA, RHIC-BES, and high-luminosity LHC runs will further constrain QCD matter under extreme conditions and establish dileptons as penetrating probes of the strongly interacting medium.


\end{document}